\documentclass[aps, prb,reprint,superscriptaddress]{revtex4-2}
\usepackage{lingmacros}
\usepackage{tree-dvips}
\usepackage{graphics}
\usepackage{graphicx}
\usepackage{epsfig}
\usepackage{epsf,epic}
\usepackage{color}
\usepackage{mathrsfs}
\usepackage{hyperref}
\hypersetup{linktocpage,colorlinks=true,allcolors=blue}
\usepackage{marvosym }
\usepackage{siunitx}
\usepackage{amsmath}
\usepackage{amssymb}
\usepackage{amsfonts}
\usepackage{wrapfig}
\usepackage{pstricks}
\usepackage{multirow}
\usepackage{pst-node}
\usepackage{extdash}
\usepackage{braket}
\usepackage{bm}
\usepackage{CJKutf8}
\usepackage{footnote}
\usepackage{dcolumn}
\usepackage{indentfirst}
\usepackage{graphicx}
\graphicspath{{figs/}}
\usepackage{upgreek}
\usepackage{makecell}
\usepackage{subfigure}
\newcommand{\etal}{\textit{et al.\ }}

\newcommand{\tc}{\textcolor}

\begin{document}
\title{Contrasting roles of the superexchange and direct exchange interactions in the metal-insulator transitions of the 2D ferromagnetic $1T$-Fe$X_2$ ($X$=Cl, Br, I)}
\author{Hongxu Luo}
\thanks{These two authors contributed equally to this work.}
\affiliation{School of Physics and Optoelectronics, Shandong Normal University, Jinan 250358, China}
\author{Baoyang Zhou}
\thanks{These two authors contributed equally to this work.}
\affiliation{School of Physics and Optoelectronics, Shandong Normal University, Jinan 250358, China}

\author{Sai Lyu}
\email{Corresponding author: sailyu@sdnu.edu.cn} 
\affiliation{School of Physics and Optoelectronics, Shandong Normal University, Jinan 250358, China}

    \begin{abstract}
\tc{black}{Abstract: Within the rapidly growing  family of two-dimensional (2D) materials, 2D ferromagnetic (FM) materials have attracted a great deal of recent attention. The ferromagnetism in 2D FM materials originates from the  underlying exchange interactions.  In particular, metal-insulator transitions (MITs) in 2D FM materials induced via various modulation strategies is critical to the design of 2D electronic devices with reduced sizes (e.g. transistor). However, the critical behaviors of exchange interactions and their effects on the Curie temperatures near MITs remains to be clarified. By taking the 2D FM $1T$-Fe$X_2$ ($X$=Cl, Br, I) monolayers as the testbeds, we computationally study the critical behaviors of the exchange interactions in the vicinity of the MITs driven by the on-site Coulomb electron-electon interactions. The FM superexchange interactions and the direct exchange interactions are found to energetically favor the halfmetallic and semiconducting states, respectively.  In addition, the halfmetallic states have the stronger FM couplings and thus the higher Curie temperatures. This study provides a comprehensive understanding of the interrelations between MITs and exchange interactions and could be useful for the device design requiring both MITs and exchange-interaction-related magnetic properties, particularly robust magnetic ordering and higher Curie temperatures, in 2D FM materials. }

\end{abstract}
\maketitle

\section{Introduction}
\tc{black}{The family of the atomically thin two-dimensional (2D) materials, such as transition-metal dichalcogenides and van der Waals ferroelectrics, has been greatly expanded since the discovery of graphene in 2007 \cite{Geim2007, Manzeli2017, Li2021, Wang2023naturemat}. The experimental realizations of 2D ferromagnetic (FM) materials were successfully demonstrated in Cr$_2$Ge$_2$Te$_6$ and CrI$_3$ in 2017 \cite{Gong2017, Huang2017, Samarth2017}, which then stimulated a great number of  experimental and theoretical studies to explore the field of 2D magnetism \cite{Burch2018, Dupont2021, Mitra2023, Kure2022, Gibertini2019, Gong2019, Mak2019}. 2D FM materials have promising technological applications for the next-generation electronics, such as low-power spin field-effect transistors \cite{Gong2019, Gibertini2019, Liu2020}. Most importantly, they can provide platforms for the fundamental studies of magnetism in the regime of the true 2D limit \cite{Mak2019, Dupont2021, Wilson2021, Mitra2023}.   }

\tc{black}{For 2D FM materials, the ferromagnetism originates from the magnetic exchange interactions between neighboring transition metal ions which mainly includes direct exchange and superexchange interaction, etc. \cite{exchange2023rmp, exchange1953rmp, Jiang2021}. Within the Heitler-London model for direct exchange interactions, two orbitals with finite overlap integrals favor the antiferromagnetic (AFM) couplings \cite{Kolos1989}. It is rather usual that FM superexchange interactions and AFM direct exchange interactions are competing. According to the Goodenough-Kanamori-Anderson (GKA) rules,  it is the superexchange interactions with nearly right-angle interaction paths that lead to the FM couplings between two cations with partially filled $d$ shells \cite{Goodenough1955, Kanamori1959, Anderson1959}. }

Metal-insulator transition (MIT) is a widely observed physical phenomenon in strongly correlated materials and remains as a fundamentally complicated problem in condensed matter physics \cite{Sordi2007, Imada1998, bookmit}.  The realization of MITs in various condensed matter systems involving the fluctuations in  the spin, charge, orbitals and lattice degrees of freedom are useful for device application, such as the transistor \cite{Imada1998, bookmit}. The emergence of 2D FM materials
has renewed interests in this respect in order to incorporate these 2D materials into device-like architectures \cite{Lowe2024, Ma2025, Gubo2024, Huangb2022, Shin2023, Gao2021sa, liuqh2021, Wu2024apl}.

For 2D FM materials, the MITs are generally accompanied by the changes in magnetic properties and/or magnetic ordering. For instance, MITs in Cr$_2$Ge$_2$Te$_6$, CrSiTe$_3$ flakes, and VI$_3$ are experimentally observed to be accompanied by considerable changes in the Curie temperatures \cite{Bhoi2021, Sun2018, Lin2018, Zhang2021nl,Valenta2021}. However, many recent studies focus on expanding the family of the 2D FM materials (e.g., 2D metal-organic frameworks) in which MITs can be induced through various controlling methods \cite{Lowe2024, Ma2025, Gubo2024, Huangb2022, Shin2023, Gao2021sa, liuqh2021, Wu2024apl}, without yet clarifying the interrelations between MITs and exchange interactions (i.e., superexchange and direct exchange) underlying the technologically important  magnetic properties and magnetic ordering, particularly robust magnetic ordering and higher Curie temperatures. 
Therefore, the critical behaviors of the exchange interactions near the Mott MITs remains as fundamental questions to be explored.  

As Mott insulators, bulk Fe$X_2$ ($X$=Cl, Br, I) materials belonging to layered transition-metal dihalids are good model magnetic systems for experimentally studying the exchange couplings, magnetic transitions, and metal-insulator transitions \cite{Rozenberg2009, Pasternak2001, Kune2003, Binek2000, Katori1996, Youn2002}. The monolayer FeCl$_2$ and FeBr$_2$ have recently been experimentally synthesized from the bulk counterparts \cite{Cai2020,  Zhou2020jpcc, Xiang2024}, which enables the FeCl$_2$ and FeBr$_2$ monolayers to be  promising candidates for studying the interrelations between exchange interactions and metal-insulator transitions down to the 2D limit \cite{liuqh2021}.

In this work, we  computationally study the fundamental interrelations between the exchange interactions and the  Mott  metal-insulator transitions  in the FM $1T$-Fe$X_2$ ($X$=Cl, Br, I) monolayers. As functions of the Hubbard $U$ parameter, the quantitative strength indicators of the superexchange interactions and the direct exchange interactions are defined to describe and interpret the critical behaviors of the strength of the exchange interactions near the Mott MITs.  We find that the FM superexchange interactions and the direct exchange interactions energetically favor the halfmetallic (HM) and semiconducting (SC) states, respectively. In addition, MITs are accompanied by enhanced FM couplings and thus Curie temperatures when crossover from SC to HM states. This study is useful for device applications which have specific requirements in the realizations of MITs and  exchange-interaction-related magnetic properties (e.g., high Curie temperatures) for 2D FM materials.

\section{Computational Methods}
\tc{black}{The first-principles calculations are performed by using the Quantum ESPRESSO \cite{qe} package. The structural relaxations and the electronic structures calculations are within the framework of the spin-polarized density functional theory (DFT) \cite{KohnSham}. We choose the generalized gradient approximation (GGA) functional developed by Perdew, Burke, and Ernzerhof \cite{pbegga} as the exchange-correlation functional and the correlations effects pertaining to the transition metal atoms are accounted for by applying the mean-field Hubbard $U$ terms to represent the on-site Coulomb repulsion energy \cite{dftu, linearldau, Pakdel2025}. We adopt the method of dispersion correction developed by Grimme \etal \cite{vdw} for the Van der Waals correction.  The GBRV ultrasoft pseudopotentials \cite{gbrvpsp} are used in the DFT calculations. The Monkhorst-Pack $\bf k$-point grid \cite{mp1976} is set as 24$\times$24$\times$1 and the plane-wave energy cutoff is 80 Ry. To minimize the spurious interactions between the periodic slabs under the periodic boundary condition, the thickness of the vacuum region is as large as $\sim$ 15 $\rm \AA$.  For each FM $1T$-Fe$X_2$ ($X$=Cl, Br, I) monolayer, the HM and SC states are obtained from the self-consistent field calculations in which the starting occupations are initialized according to the corresponding electronic band structures \cite{liuqh2021}. To quantitatively describe and interpret the critical behaviors of the exchange interactions near the MITs,  the strength indicators of the superexchange interactions and the direct exchange interactions are defined \cite{Murrell1970, Lyu2023prb}. According to the classical Heisenberg exchange model, the parameters $J$   are determined from the total energy differences between the FM and the  corresponding AFM states for each $1T$-Fe$X_2$ ($X$=Cl, Br, I) monolayer \cite{Kulish2017, Botana2019}. }

\section{Results and Discussion}
\subsection{Electronic band structures and Mott metal-insulator transitions}

\begin{figure}[tbp]
\hspace*{-0.2cm}%
\centering 
\includegraphics[width=4.5 cm]{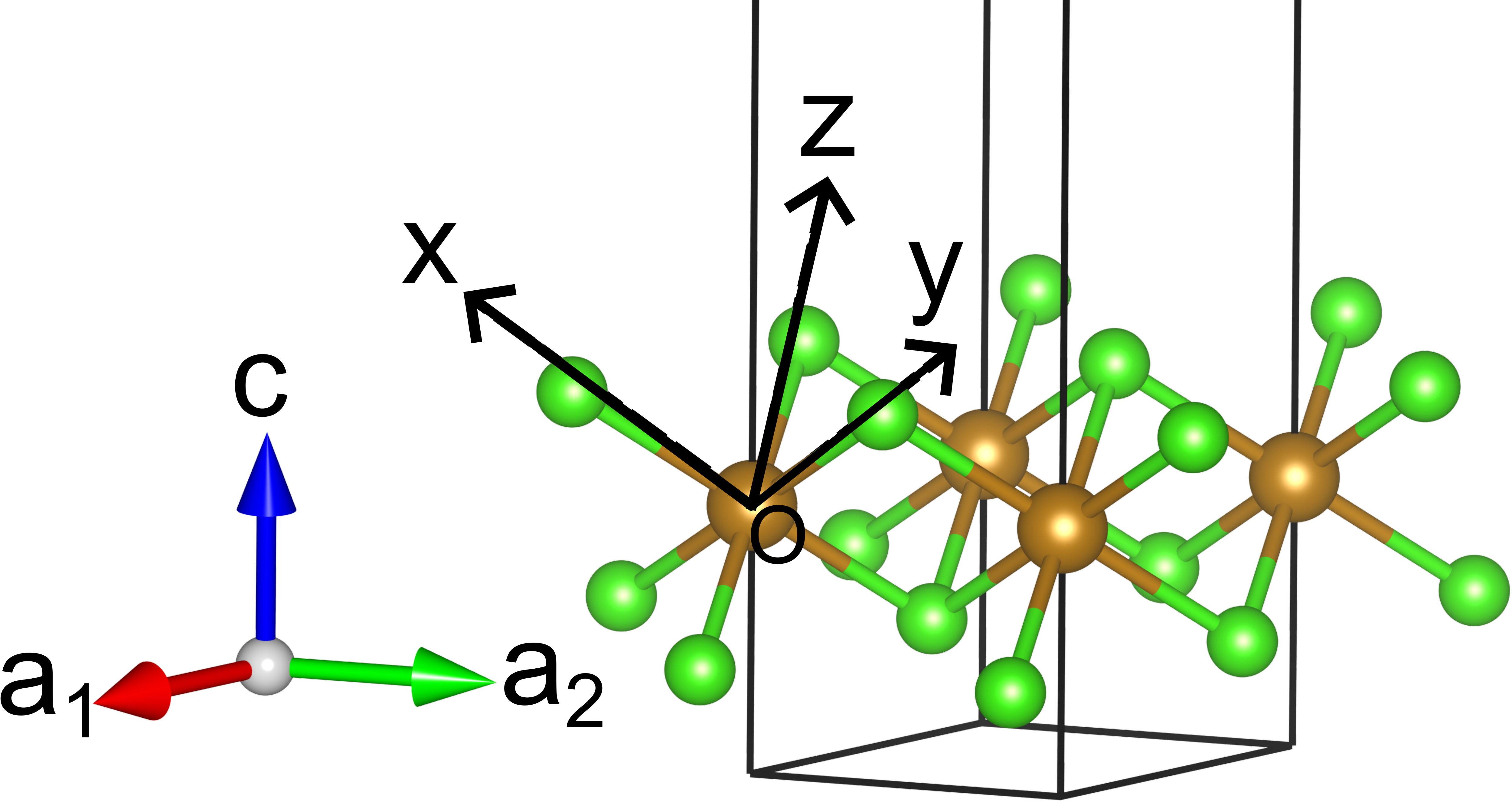}  
\includegraphics[width=3.5 cm]{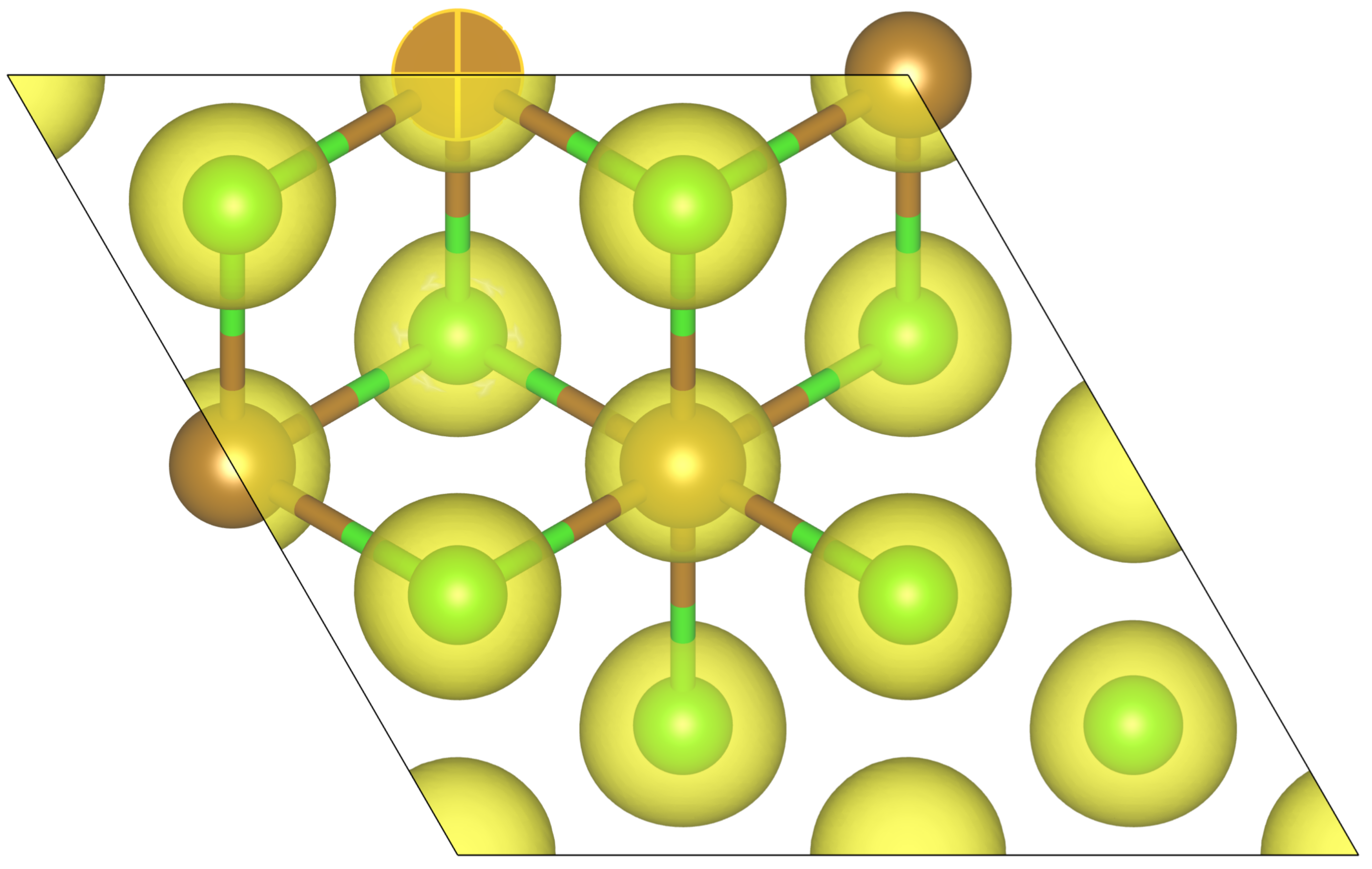}  \\  \vskip -0.25 cm

     	\caption{\tc{black}{Crystal structure and the electron density plot of the two-dimensional $1T$-Fe$X_2$ ($X$=Cl, Br, I) monolayers. The Fe and $X$ atoms are denoted by the brown and green spheres, respectively. The lattice vectors (i.e., $\vec{a_1}$, $\vec{a_2}$, and $\vec{c}$) and the Cartesian axes (i.e., $Oxyz$) are schematically shown. }   \label{figstr}}   
    \end{figure} 

The crystal structures of the $1T$-FeX$_2$ ($X$=Cl, Br, I) monolayers are shown in Fig.\ \ref{figstr}. These crystal structure are visualized by the VESTA software \cite{vesta}. The lattice is hexagonal and the space group is $P\bar{3}m1$ (No. 164). Each Fe layer is sandwiched by two halogen layers.  Six valence electrons remain in the Fe-$3d$ orbitals and the corresponding spin quantum number $S$ is 2.  The obtained lattice parameter $a$ for FeCl$_2$, FeBr$_2$ and FeI$_2$ are 3.58 \AA, 3.77 \AA, and 4.05 \AA, respectively.  $c$ is set as 20 $\textrm{\AA}$ to minimize the spurious interactions between the periodic slabs. Each central cation and the neighboring 6 anions forming an octahedron. The octahedral crystal field splits the five $d$ orbitals into the two sets, i.e., $e_g$ and $t_{2g}$. There are two degenerate energy-higher $e_g$ atomic orbitals (i.e., $d_{z^2}$ and $d_{x^2-y^2}$) and three degenerate energy-lower atomic $t_{2g}$ orbitals (i.e., $d_{xy}$, $d_{yz}$ and $d_{zx}$). 
When forming the electronic bands, the distorted octahedral crystal field split degenerate $t_{2g}$ orbitals into an $e_g^*$ doublet and an $a_{1g}$ singlet. The corresponding wavefunctions can be written as \cite{Botana2019}
\begin{eqnarray}
|e_{g\pm}^{*}\rangle &=& \pm\frac{1}{\sqrt{3}}(|d_{xy}\rangle+ e^{\mp 2 \pi i/3}| d_{yz}\rangle+e^{\pm 2 \pi i/3}|d_{zx}\rangle), \nonumber \\
|a_{1g}\rangle &=& \frac{1}{\sqrt{3}}(| d_{xy} \rangle+| d_{yz}\rangle+|d_{zx} \rangle).
\end{eqnarray}

\begin{figure}[t]
\hspace*{-0.2cm}%
\centering 
\subfigure[]{\includegraphics[width=4 cm]{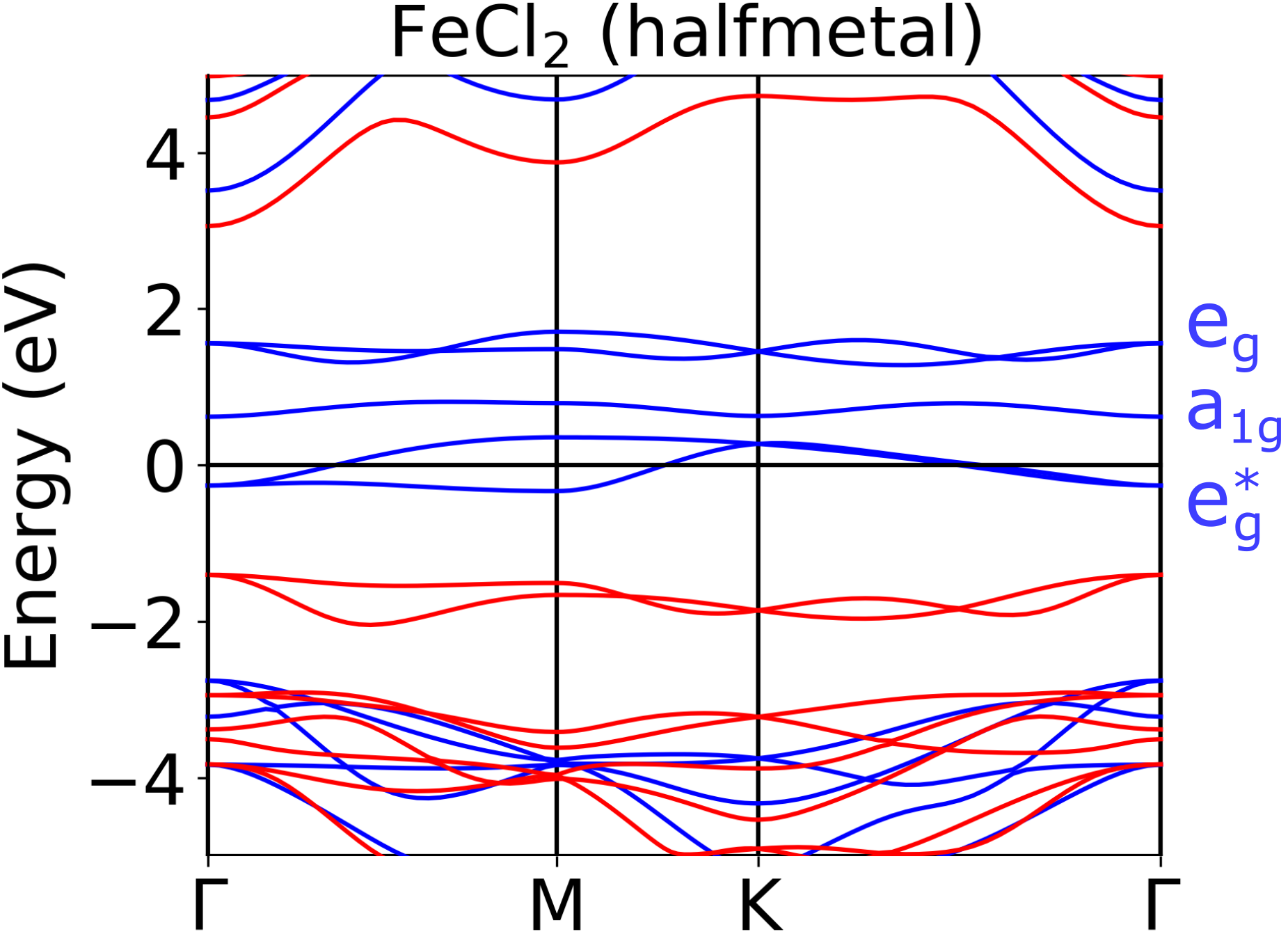}  }
\subfigure[]{\includegraphics[width=4 cm]{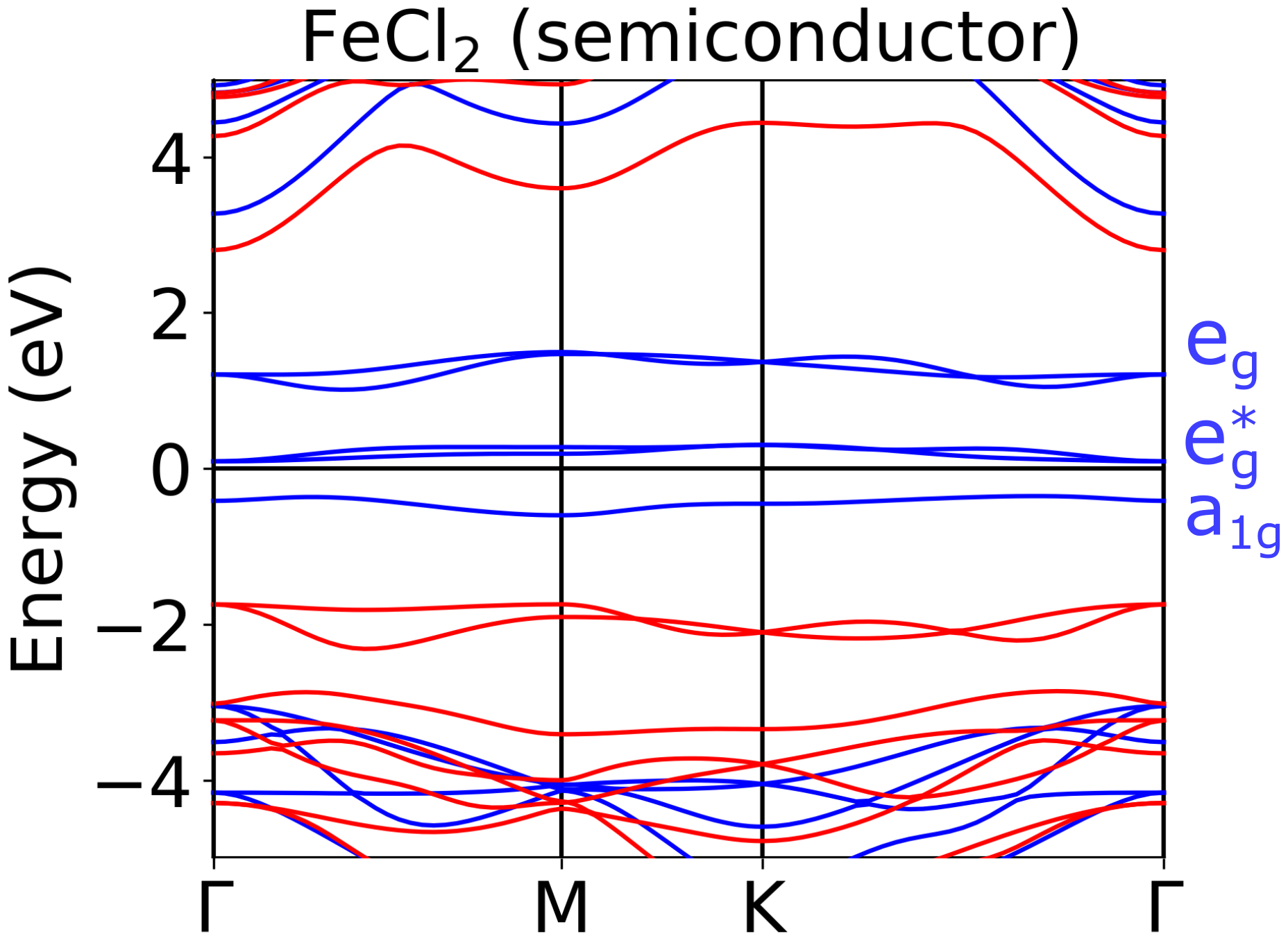}  }\\  
     	\caption{Electronic band structures of the ferromagnetic (a) halfmetallic and (b) semiconducting $1T$-FeCl$_2$. \label{figband}}   
    \end{figure} 

In Fig.\ \ref{figband}, we show the  electronic band structures for the FM $1T$-FeCl$_2$ monolayer in which the band labels are indicated. The Hubbard $U$ parameter is 0.67 eV. The band structures and projected density of states (PDOS) for the $1T$-Fe$X_2$ ($X$=Cl, Br, I) monolayers are provided in Figs. S1 and S2 in the Supplementary material.  In the FM HM states, the band ordering in the five spin down $d$ bands near the Fermi levels are $e_g > a_{1g} >e_g^*$ in terms of  the Kohn-Sham (KS) eigenvalues. The lowest two $e_g^*$ bands are occupied by a spin-down $d$ electron and are crossing with each other. As a result, both of the two $e_g^*$ bands are partially filled. At variance, the band ordering is $e_g > e_g^* > a_{1g}  $ for the FM SC states. The $a_{1g}$ band is completely occupied by the remaining one electron in the spin down $d$ channel. Clearly, the FM HM and FM SC states have the contrasting ordering of $e_g^{*}$ and $a_{1g}$ bands. For FM SC and HM states, the ordering of five $d$ bands near the Fermi levels remain unchanged when the Hubbard $U$ parameter is varied.


\begin{figure}[!htbp]
\hspace*{-0.2cm}%
\centering 
\subfigure[]{\includegraphics[width=4 cm]{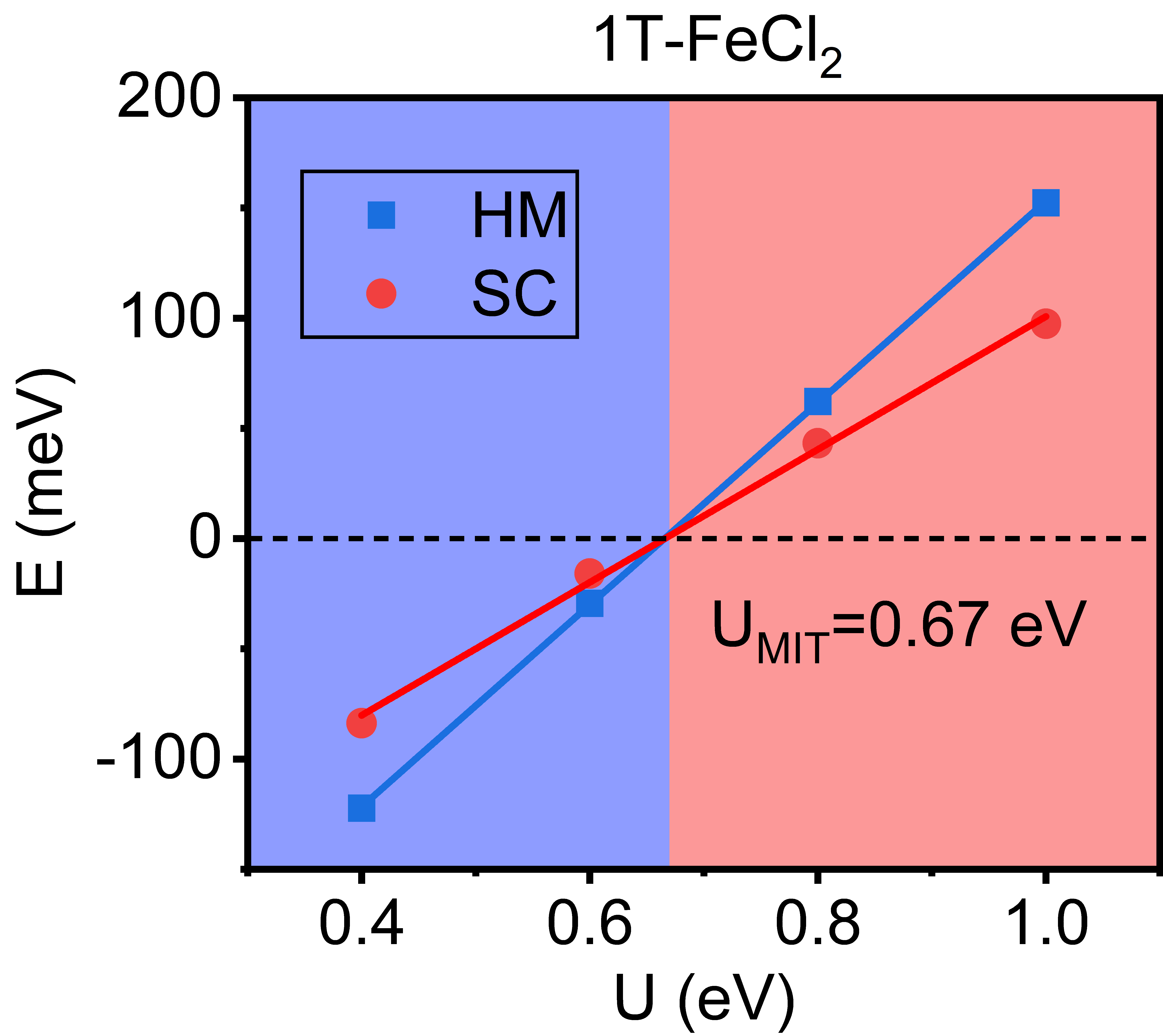}  }
\subfigure[]{\includegraphics[width=4 cm]{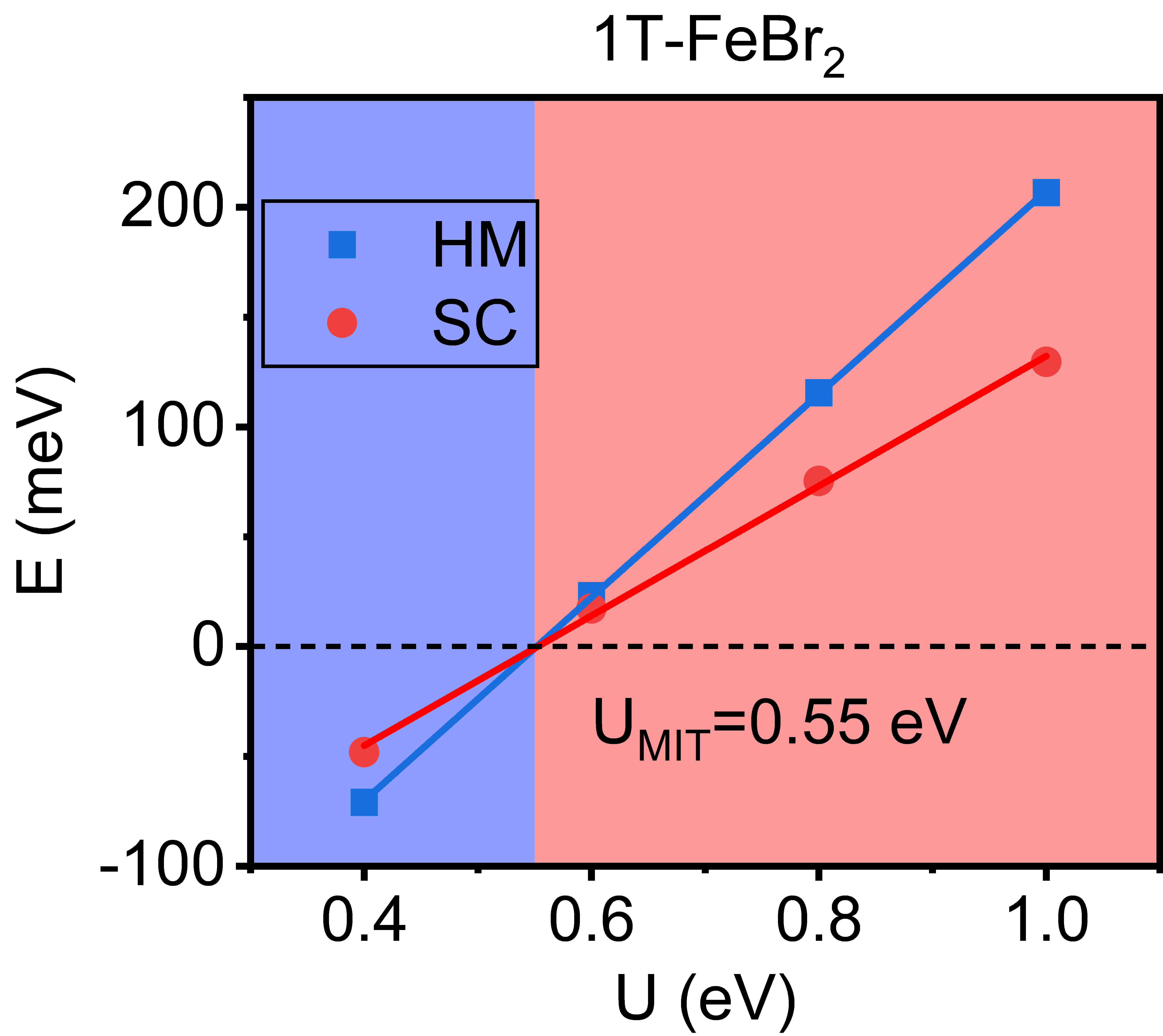}  }  
\subfigure[]{\includegraphics[width=4 cm]{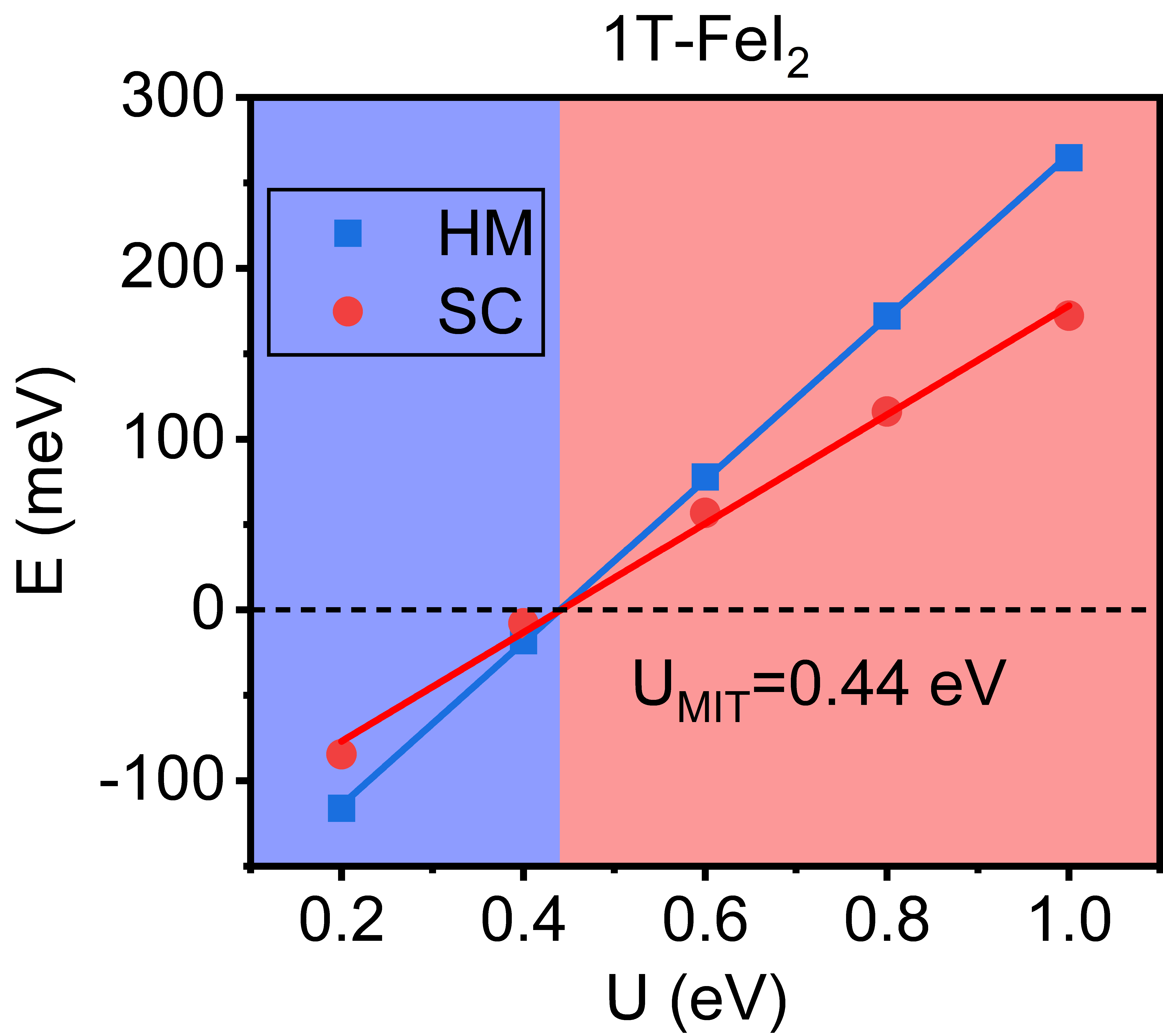}  }\\  \vskip -0.25 cm

     	\caption{Total energies per unit cell as functions of the Hubbard parameter $U$ for (a) FeCl$_2$, (b) FeBr$_2$, and (c) FeI$_2$.  For each monolayer, the total energies at $U_{\textrm{MIT}}$ are taken as the references. \label{figmit}}   
    \end{figure}

\tc{black}{The phase diagram of metal-insulator transitions in materials are usually interpreted by the electron correlation strength, which can be represented by U/W, where $U$ represents the on-site Coulomb interaction energy and can drive MITs \cite{Imada1998}. For the $1T$-Fe$X_2$ ($X$=Cl, Br, I) monolayers, the total energies per unit cell of the FM HM and FM SC states as functions of the Hubbard $U$ parameters are shown in Fig.\ \ref{figmit}. The critical $U$ parameters are denoted as $U_{\textrm{MIT}}$ where the total energies of the FM and FM SC states crossover. The total energies at $U_{\textrm{MIT}}$  are taken as the references. The values of  $U_{\textrm{MIT}}$ display an decreasing trend from FeCl$_2$ to FeBr$_2$ to FeI$_2$. When $U < U_{\textrm{MIT}}$, the FM state has lower energy. At variance, the FM SC state has lower total energy when $U > U_{\textrm{MIT}}$. This means stronger electronic correlation energetically prefers to the FM SC states rather than the FM HM states and the MITs should be classified as the Mott type. }

\subsection{Critical behaviors of the superexchange and direct exchange interactions}

\begin{figure}[t]
\hspace*{-0.2cm}%
\centering 
\subfigure[]{\includegraphics[width=3.6 cm]{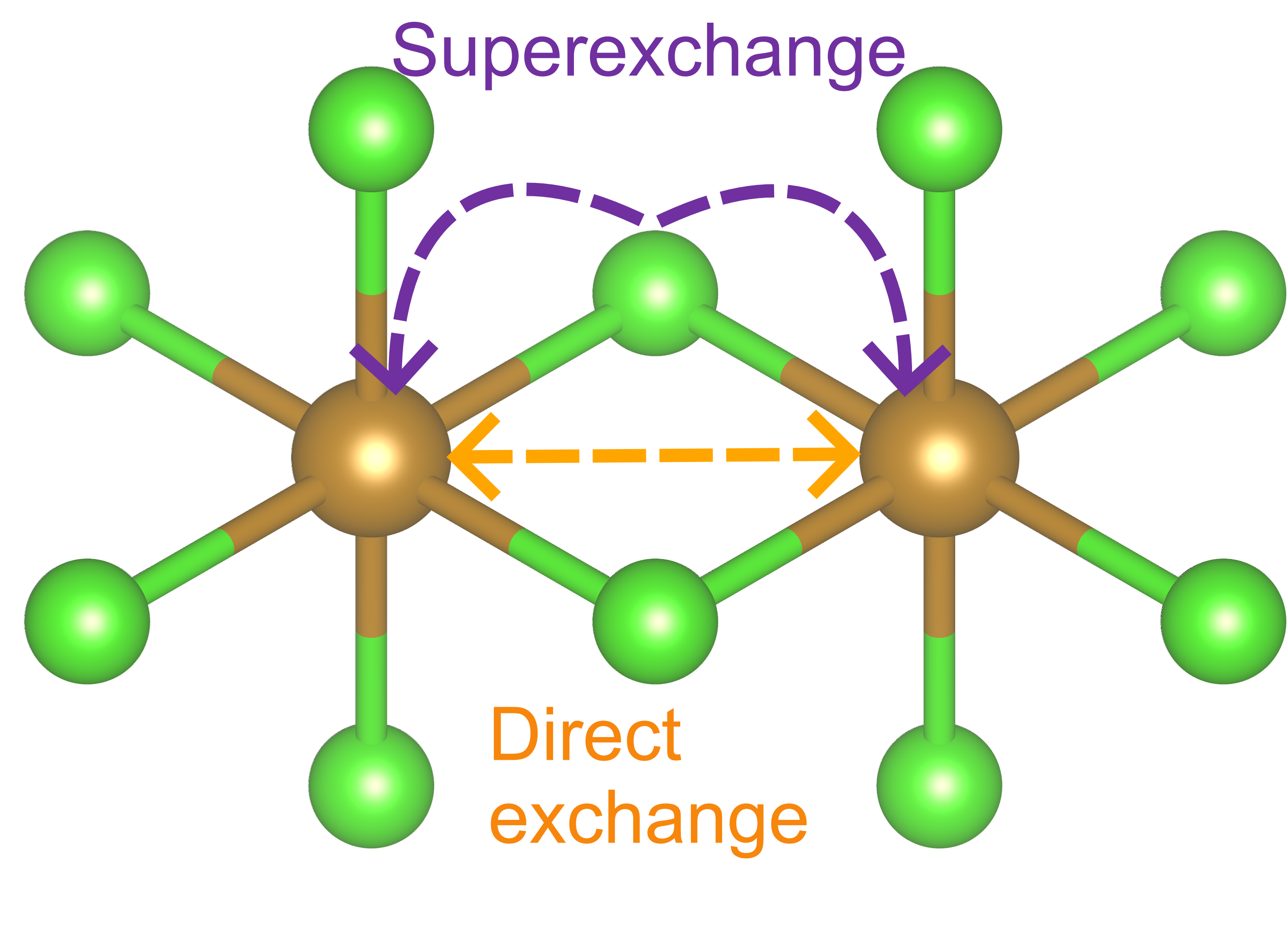} } 
\subfigure[]{\includegraphics[width=3.6 cm]{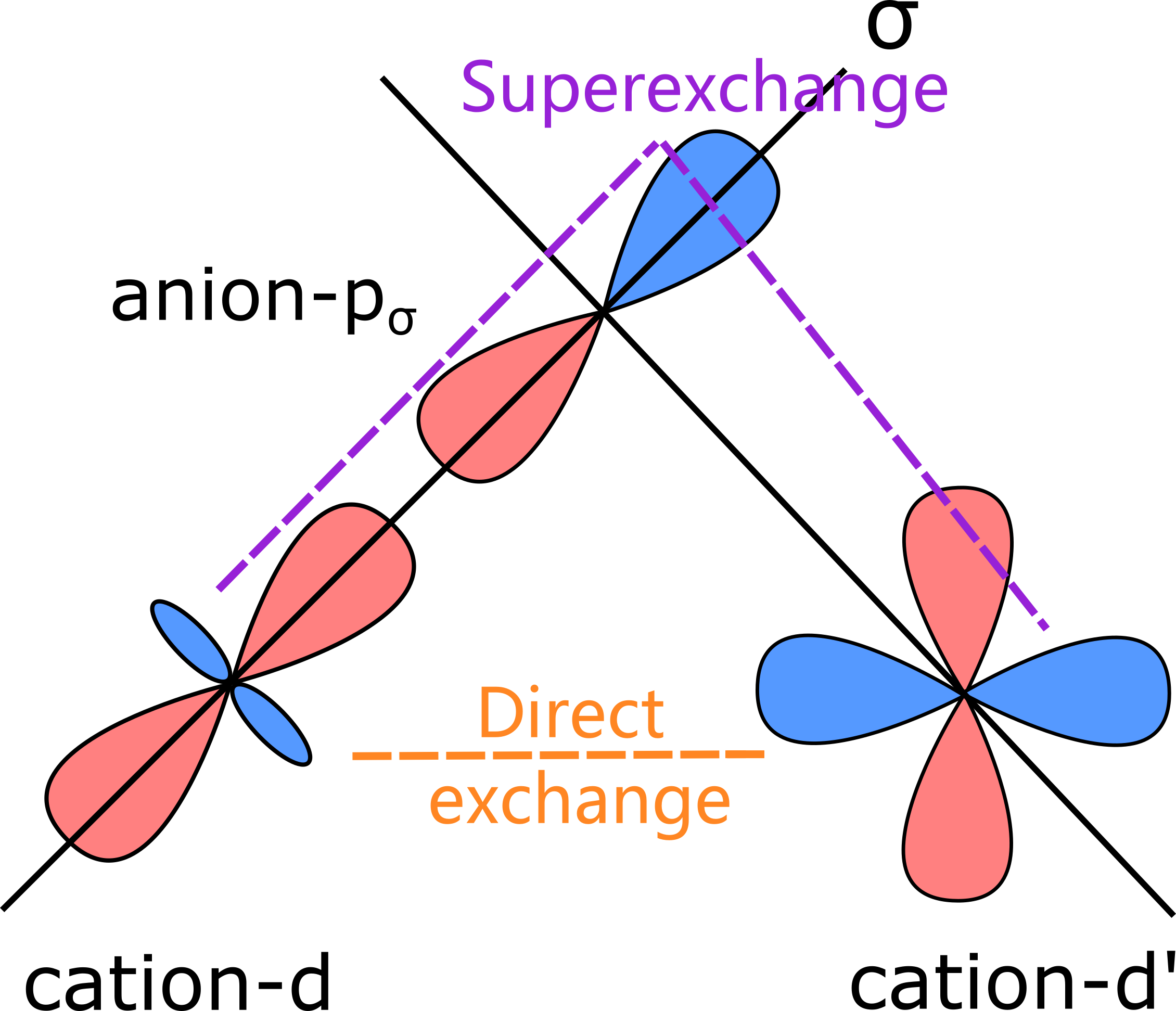} }  \\
\subfigure[]{\includegraphics[width=6.4 cm]{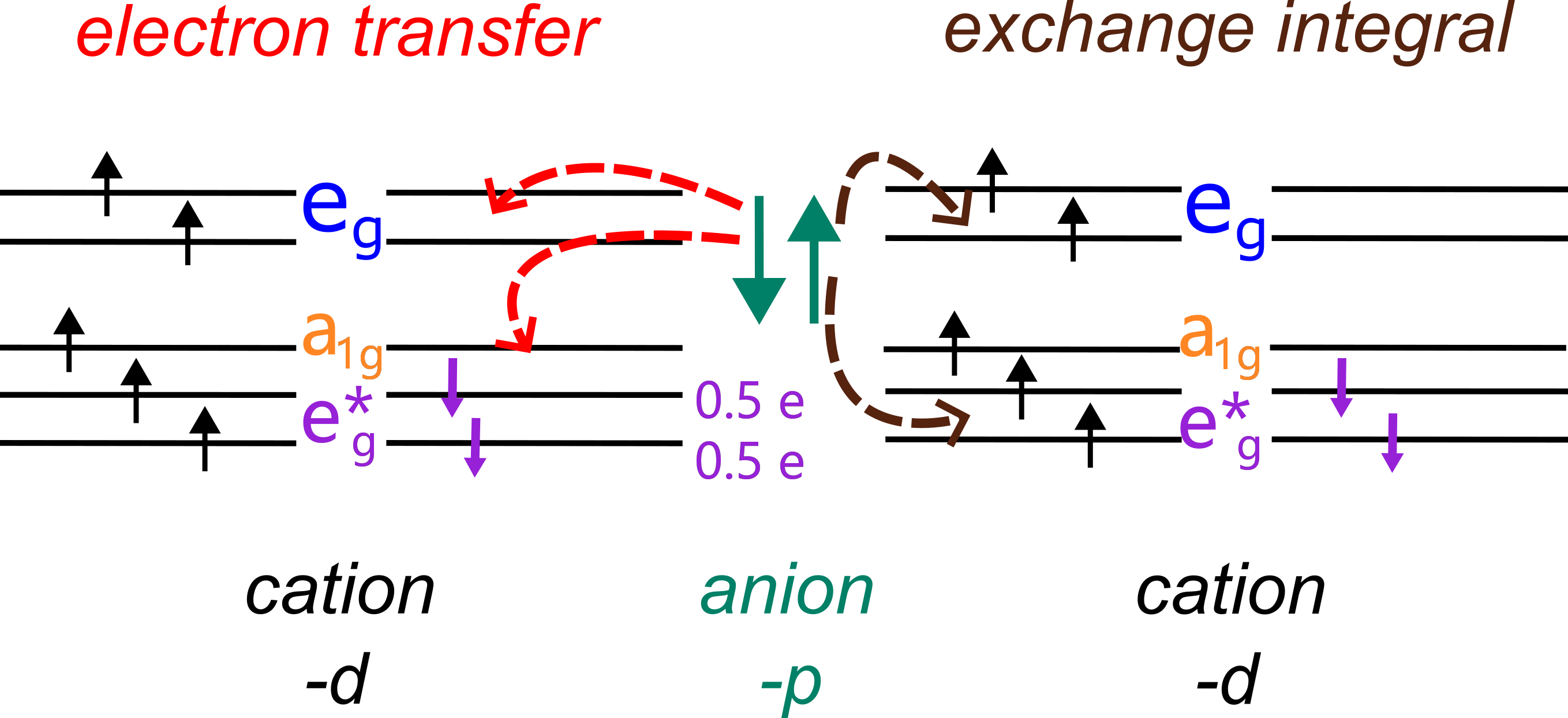} }
\subfigure[]{\includegraphics[width=6.4 cm]{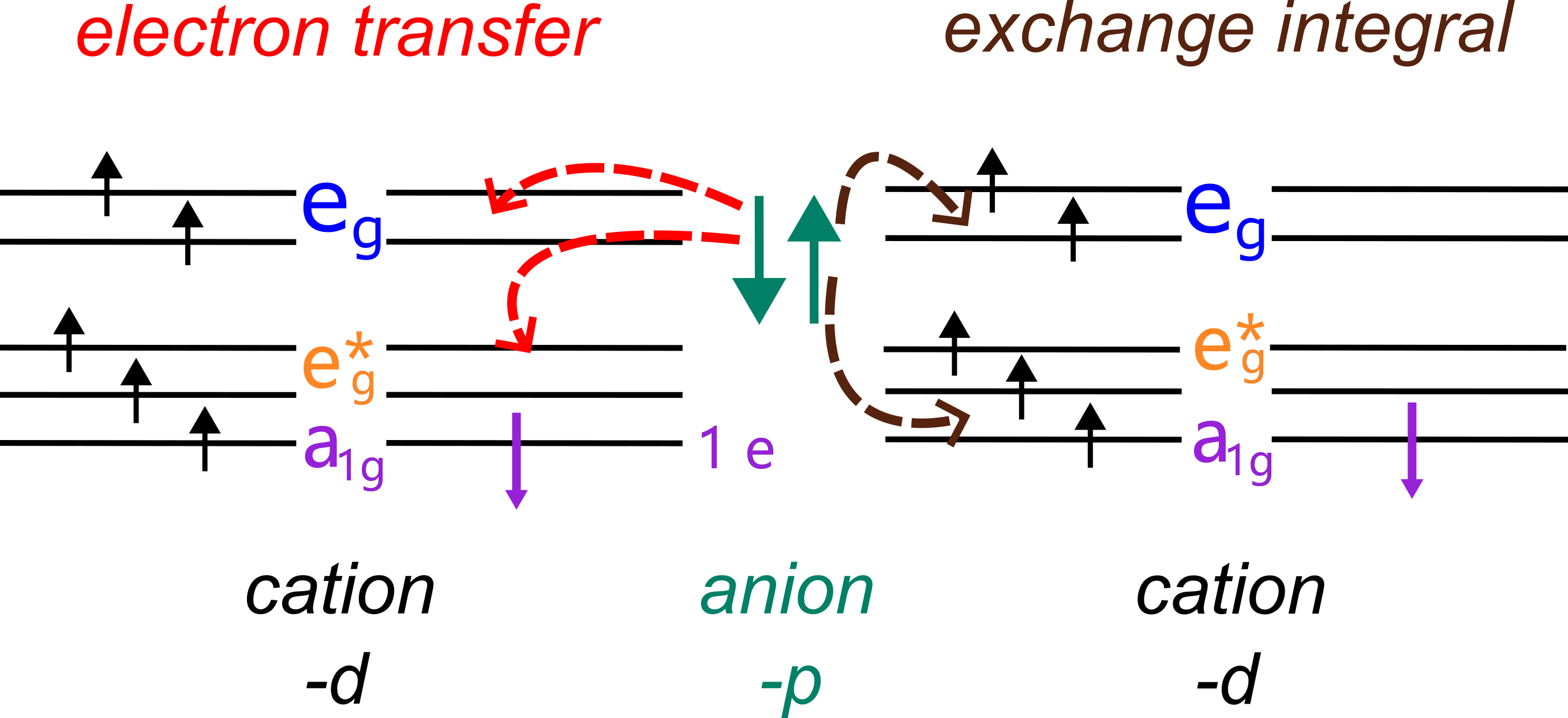} } \vskip -0.25 cm
     	\caption{(a) Interaction path, (b) the involved atomic orbitals  for the exchange interactions, the mechanism for the FM superexchange interactions in the (c) halfmetallic and (d) semiconducting $1T$-Fe$X_2$ ($X$=Cl, Br, I) monolayers. \label{figsuper}}   
    \end{figure}

In FM $1T$-Fe$X_2$ ($X$=Cl, Br, I) monolayers, the direct exchange interactions between the neighboring Fe cations results in the AFM couplings. In contrast, superexchange interactions between neighboring Fe cations involve intervened ligand atoms and $\sim 90 ^ \circ$ superexchange interaction paths result in the FM couplings (cf. Fig.\ \ref{figsuper} (a)-(b)). Thus, the FM superexchange interactions counteract with the AFM direct exchange interactions. The FM superexchange interactions and the AFM direct exchange interactions stabilize and destabilize the FM states, respectively. As a result, the  direct exchange interactions can elevate the total energies for the FM states, whereas  the superexchange interactions conversely decrease the total energies.



The magnetic moment of the Fe atom and the total energies are only slightly different between the ferromagnetic insulating and metallic states. So we use the localized Heisenberg model for both the insulating and metallic states. Based on the Heisenberg exchange model, the isotropic exchange parameter $J$ can be obtained by calculating the energy differences between the FM and the AFM states.
The corresponding classical Hamiltonian is defined as
\begin{equation}
H=-\frac{1}{2} \sum_{i,j; i \neq j} J_{ij} S_iS_j,
\end{equation}
where $S_i$ is the spin operator at the cation site $i$ and $J_{ij}$ is the exchange coupling parameter between the cations at the sites $i$ and $j$. The summation runs over all the neighboring cation sites. 
Within a $2\times1\times1$ supercell, the two Fe cations have the same and opposite directions of magnetic  moment for the FM and AFM configurations, respectively. 
Considering the exchange couplings between the first nearest-neighboring cation sites \cite{Kulish2017, Botana2019, Li2022}, the $J$ parameters can be extracted as
\begin{equation}
J=\frac{E_{\textrm{AFM}}-E_\textrm{FM}}{8S^2}
\end{equation}
The corresponding cation-cation distance is the lattice parameter $a$. The exchange parameters corresponding to other nearest-neighboring cation sites are not considered because they are significantly smaller.\cite{Kulish2017, Botana2019, Li2022}

\tc{black}{To quantitatively interpret the behaviors of the superexchange interactions near the MITs, path-resolved strength indicators for the FM superexchange interactions are defined as \cite{Lyu2025janus}: 
\begin{equation}
J^*_{\textrm{super}} =\sum_{d, p, d'} Q_{\mathrm{ET}}(d,p)  \frac{S_{\psi}^2(p, d')}{r(p, d')} 
\label{eqJind}.
\end{equation} 
The summation runs over for all the involved orbitals.
This mechanism for the superexchange interactions is based on the GKA mechanism \cite{Goodenough1955, Goodenough1963, Kanamori1959, Anderson1950, Anderson1959} and is illustrated in Fig.\ \ref{figsuper} (c) and (d) for the halfmetallic and semiconducting states, respectively. 
The quantity $Q_\mathrm{ET}$ denotes the electron transfer from the anion-$p$ to the spin-down Fe cation-$d$ orbitals, which provides a measure of the $p-d$ bonding covalency. By calculating the integral of the corresponding PDOS of the spin-$d$ orbitals between $-\infty$ and the Fermi levels, the amount of the electrons $Q$ in the spin-down cation-$d$ orbitals are obtained. The calculated values of $Q$ are provided in Table S1 in the Supplementary material.  Based on the Hund's rule, the spin-down $e_g$ orbitals are empty, whereas each spin-down $t_{2g}$ orbital is occupied by $1/3$ electrons. For each $e_g$ orbital, $Q_\textrm{ET}$ has the same value as $Q$. For each $t_{2g}$ orbital, the corresponding $Q_{\textrm{ET}}$ is the amount of electrons more than $1/3$. The $S_{\psi}$ and $r$ denotes the overlap integrals and distance between anion-$p$ and cation-$d'$ orbitals. 
By the definition, the strength indicators $J^*_{\textrm{super}}$ have the expected orders of magnitude of the $p-d$ overlap (i.e., $J^*_{\textrm{super}} \sim S_{\psi}^4$), which is consistent with the relation obtained by considering the orbital overlapping within the Heitler-London picture \cite{Yamashita1958, Lyu2024prb2}.   }

\begin{figure}[t]
\hspace*{0.1cm}%
\centering 
\subfigure[]{\includegraphics[width=6.5 cm, height=5.2 cm]{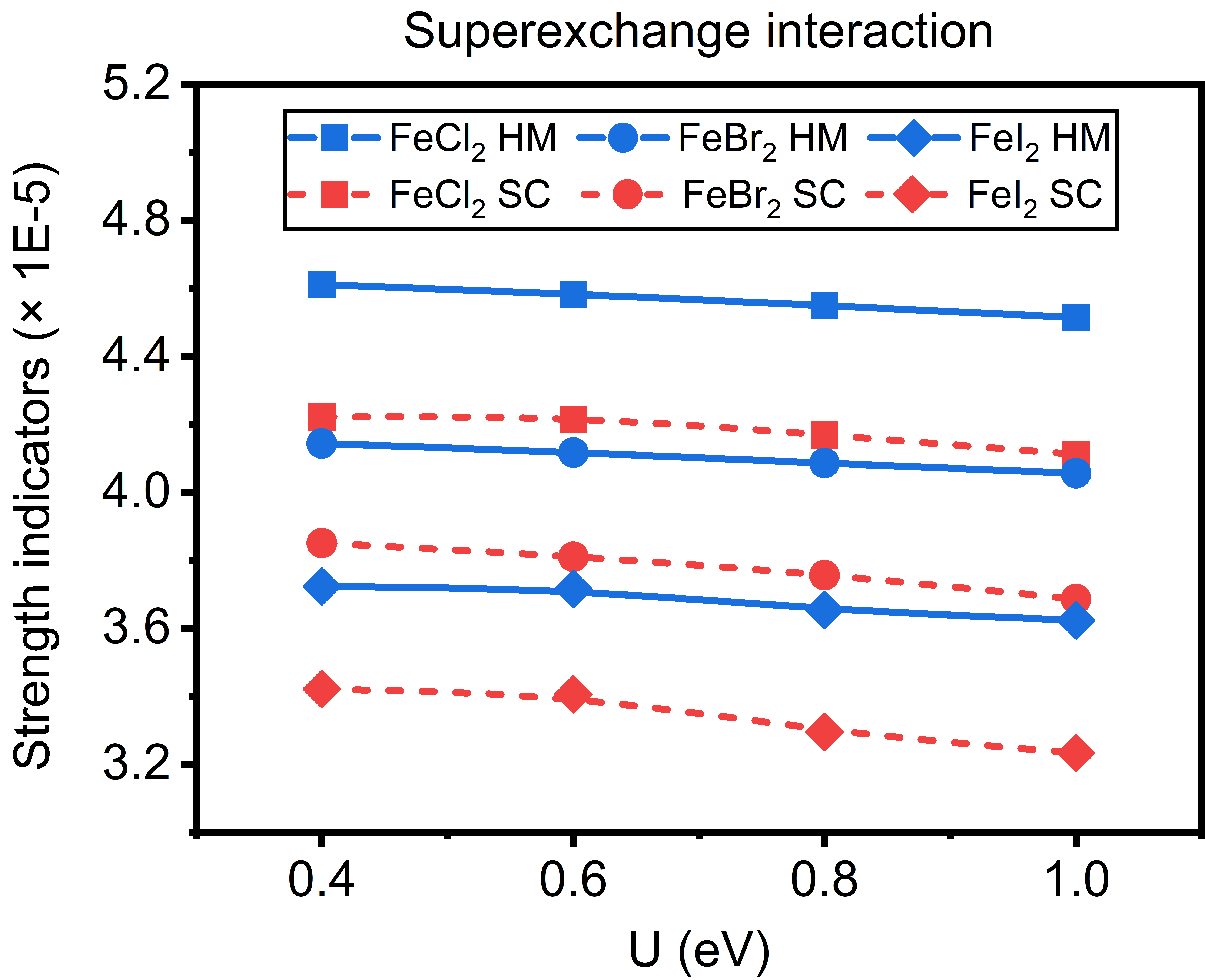}  }   
\subfigure[]{\includegraphics[width=6.8 cm, height= 5.3 cm]{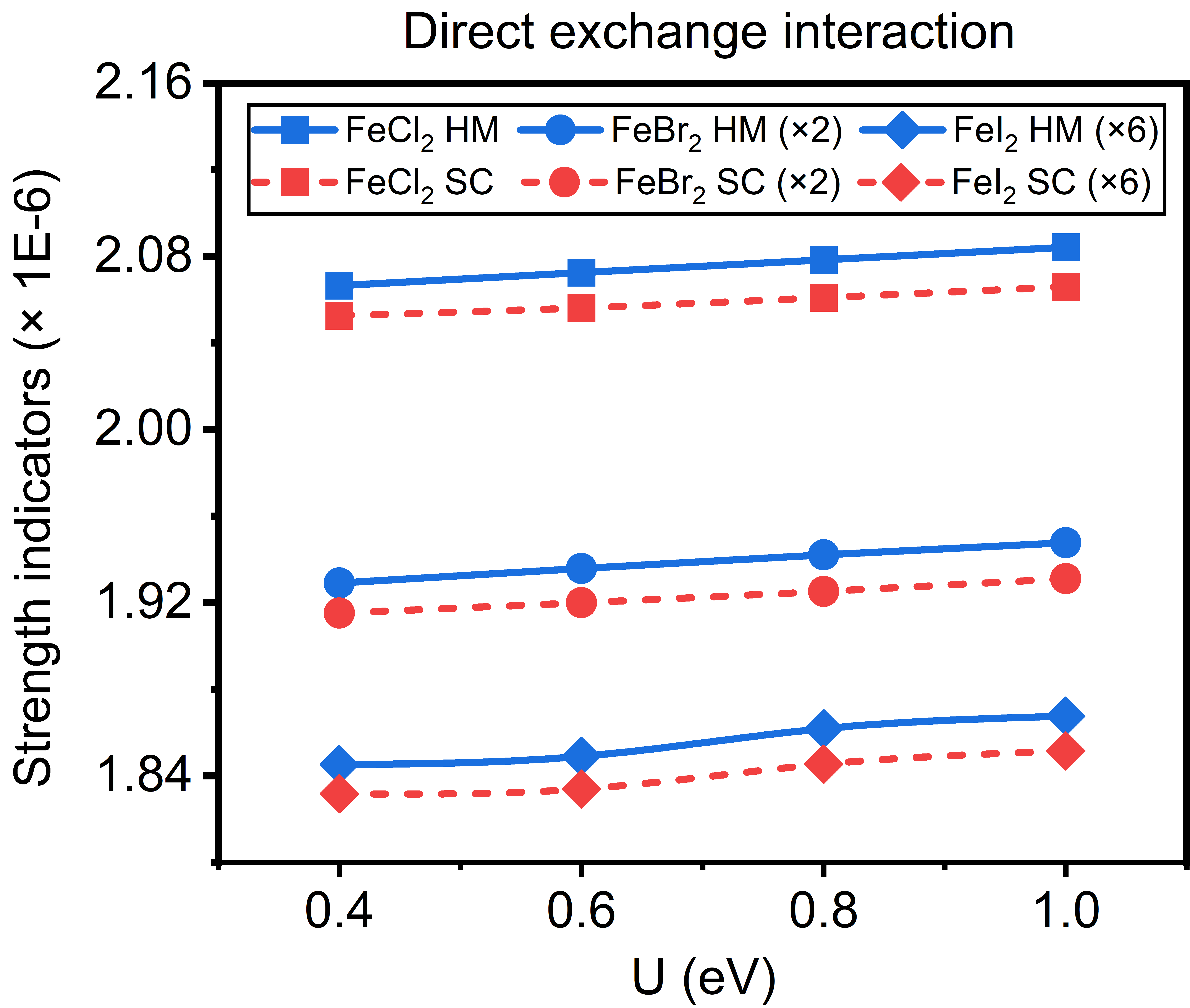}  } \\  
     	\caption{Strength indicators of  (a) superexchange interactions and (b) direct exchange interactions for the FM $1T$-Fe$X_2$ ($X$=Cl, Br, I) monolayers. \label{figufunction}}   
    \end{figure}

\tc{black}{In Fig.\ \ref{figufunction} (a), we show the calculated strength indicators of the superexchange interactions for both the FM SC and FM HM states. As the $U$ parameter increase, the strength indicators of FM superexchange interactions are decreasing. This could be attributed to the fact that the large on-site Coulomb repulsion increase the energy gaps between the cation-$d$ and the anion-$p$ orbitals. As a result, the electron transfer from anion-$p$ to the spin-down cation-$d$ orbitals and therefore the strength indicators are reduced. Most importantly, regardless the values of the $U$ parameters, the FM HM states have stronger FM superexchange interactions. This means FM superexchange interactions in Fe$X_2$ ($X$=Cl, Br, I) monolayer energetically prefer to the FM HM states rather than the FM SC states.  }

The direct exchange interactions result in AFM couplings and therefore elevate total energies in the FM states. To quantitatively study the roles of direct exchange interactions, the direct exchange energy can be approximated as $E \sim S_{\psi}^2/r$, where $S_{\psi}$ is the overlap integrals between the neighboring $d$ orbitals and $r$ denotes the cation distances \cite{Murrell1970, Lyu2025janus}. 
So we define the strength indicators for the direct exchange interactions as:
\begin{equation}
J^*_{\textrm{direct}}=\sum_{d,d'} Q(d)Q(d')  \frac{S_{\psi}^2(d,d')}{r (d,d')} ,
\label{eqJind}
\end{equation}
The summation runs over for the involved $d$ orbitals.
In Fig.\ \ref{figufunction} (b), we show the indicators of direct exchange interactions for both the FM SC and FM HM states in the $1T$-Fe$X_2$ ($X$=Cl, Br, I) monolayers. As the $U$ values increase, the on-site Coulomb repulsion get stronger and the electron transfer from the anion-$p$ to the spin-$d$ orbitals is suppressed. As a result,  the strength indicators of direct exchange interactions increase as the $U$ values increase. Clearly, the FM HM states always have stronger direct exchange interactions than the FM SC states regardless of the values of $U$. This indicates that the direct exchange interactions energetically prefer to the FM SC states rather than the FM HM states.

\begin{figure}[t]
\centering 
\subfigure[]{\includegraphics[width=6.5 cm]{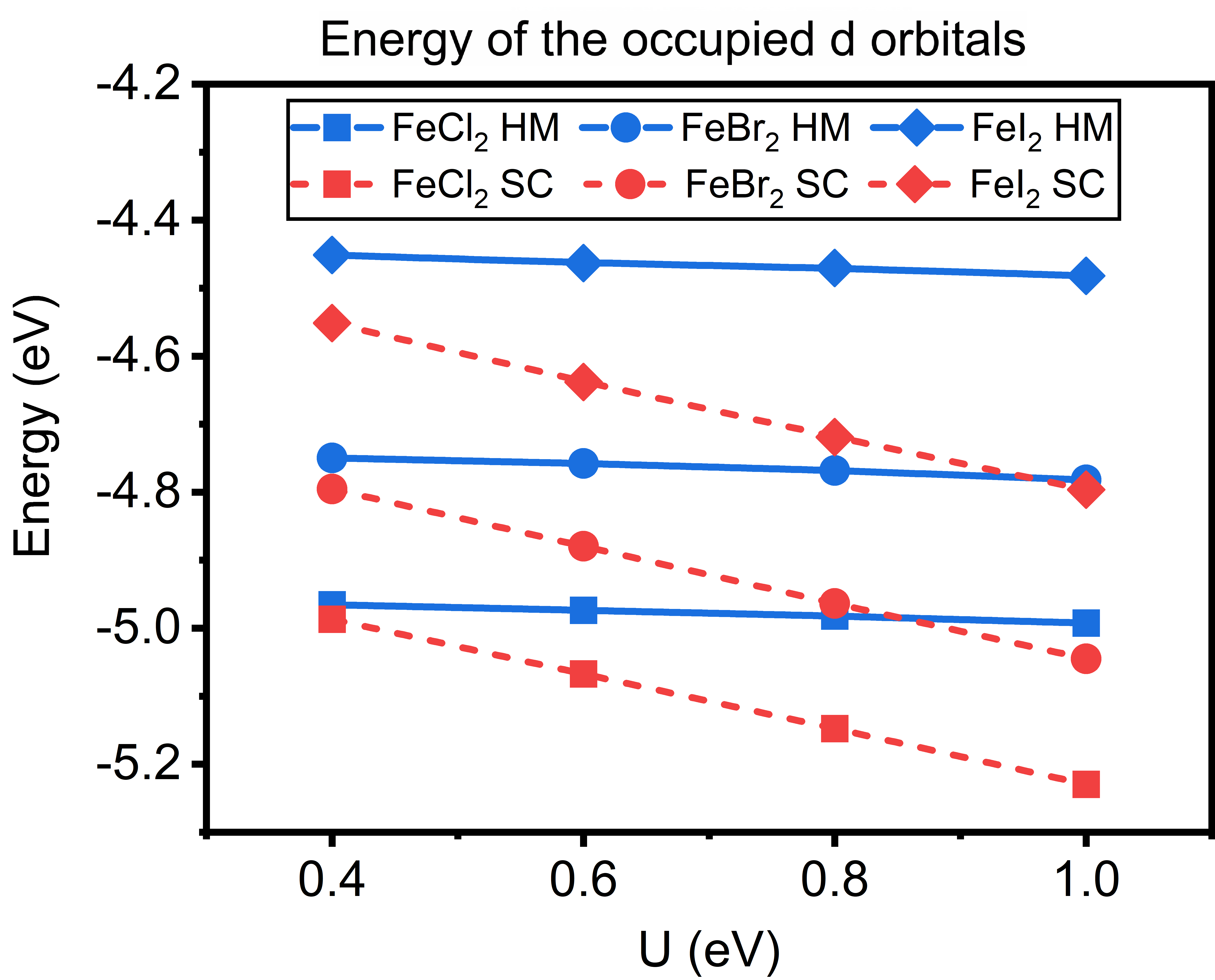}  }  
\hspace*{0.25 cm}%
\subfigure[]{\includegraphics[width= 6.25 cm, height=5 cm]{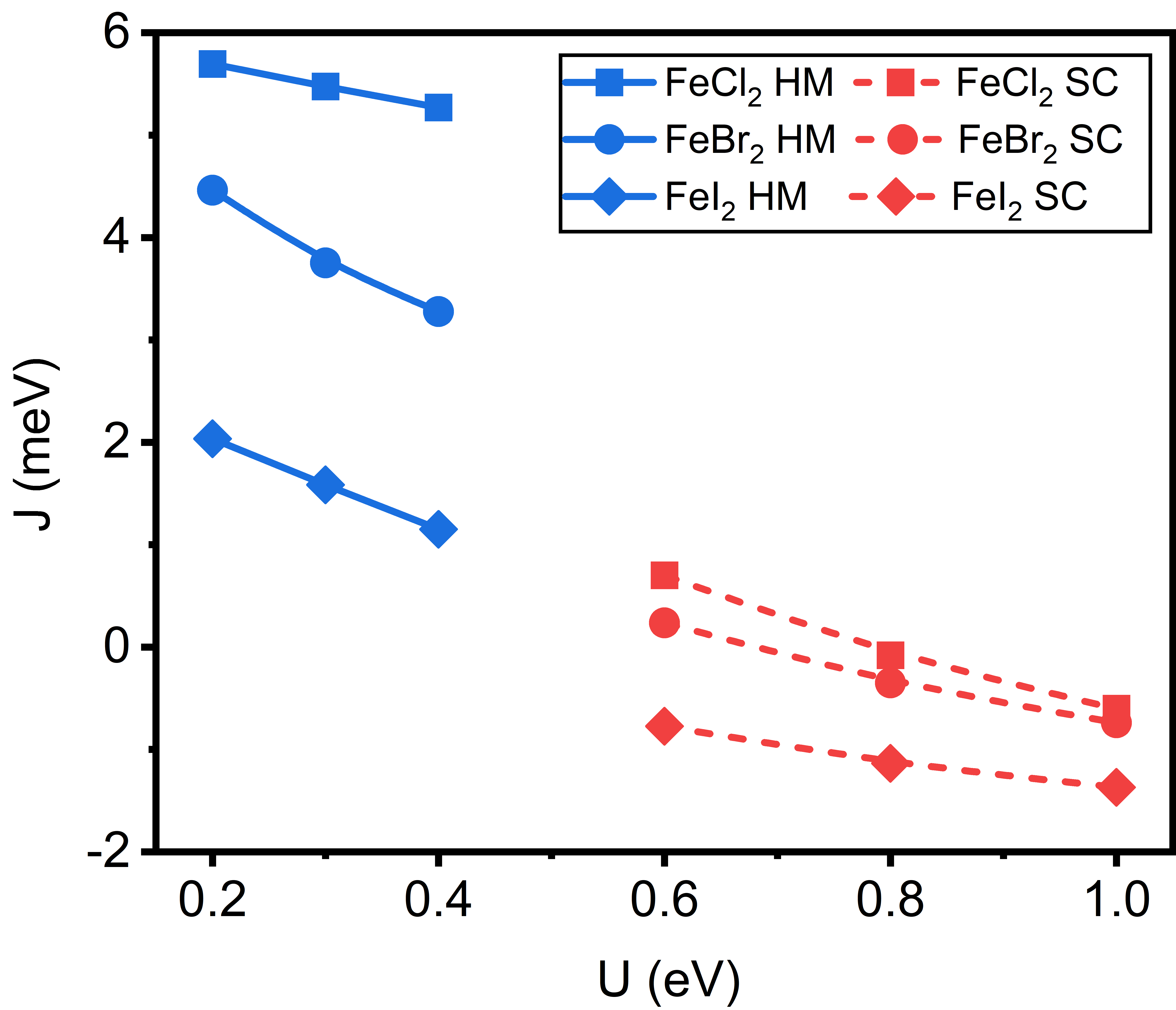}  } \\  \vskip -0.25 cm
     	\caption{(a) Total energies of the occupied spin-down $d$ bands and (b) the calculated exchange parameters $J$  for the FM $1T$-Fe$X_2$ ($X$=Cl, Br, I) monolayers. \label{figujband}}   
    \end{figure}

Exchange interactions account for the interaction energies between the neighboring occupied cation-$d$ orbitals. In addition to this, the cation-$d$ orbitals themselves contribute to the total energies. 
To quantitatively evaluate the contributions of the occupied spin-down $d$ (denoted as $d\downarrow$) orbitals to the total energies of the FM HM and FM SC states, we calculate the quantity as
\begin{equation}
\varepsilon_\textit{d}=\int^{E_\textrm{F}} \text{PDOS}_{d\downarrow}(E_{d\downarrow}) E_{d\downarrow}dE,
\end{equation}
where $E_\textrm{F}$ denotes the Fermi levels. This expression is consistent with the definition of the so-called $d$-band center \cite{Fu2020}.The energy $E$ is obtained by taking the electrostatic potential in the vacuum region $V_\textrm{vacuum}$ as the reference level. To be more specific, $E=E_{\textrm{KS}}-eV_\textrm{vacuum}$ where $E_{\textrm{KS}}$ denote the Kohn-Sham eigenvalues.  This is a common practice to obtain the so-called absolute energy levels in band offset calculations in order to compare the band edge levels pertaining to different semiconductors \cite{Lyu2024jpd, Hinuma2017}. The obtained energies of the occupied $d$ orbitals ($E_d$) are shown in Fig.\ \ref{figujband} (a). For each $1T$-Fe$X_2$ ($X$=Cl, Br, I) monolayer, the values of $E_d$  becomes lower when the $U$ values increase. This is because that larger on-site Coulomb repulsion enlarges the energy splittings between the occupied  and unoccupied $d$ orbitals. Near the Mott transition point (i.e., $U_{\textrm{MIT}}$), the FM SC state has lower $E_d$ values than the FM HM state for each monolayer. This indicates that the FM SC states are energetically favored due to the ordering of the spin-down $d$ bands.

In Fig.\ \ref{figujband} (b), we show the calculated exchange parameters $J$. Clearly, the FM HM states have larger exchange couplings than the FM SC states for each $1T$-Fe$X_2$ ($X$=Cl, Br, I) monolayer. At the Mott transition points (i.e., $U_{\textrm{MIT}}$), the positive $J$ parameters indicate that the FM superexchange interactions are more dominant than the direct exchange interactions.  As the Hubbard $U$ parameter increase, the exchange coupling parameters $J$ are decreasing for both the FM SC and FM HM states. This can be attributed to the fact that the increasing on-site Coulomb repulsion energy (i..e, the Hubbard $U$ parameter) decreases the electron transfer from anion-$p$ orbitals to the cation-$d$ orbitals and thus decreases the strength of more dominant FM superexchange interactions. 
Given that the FM HM states have slightly stronger direct exchange interactions (cf. Fig.\
 \ref{figufunction} (b)) and much larger $J$ parameters, the FM HM should have much stronger FM superexchange interactions than the FM SC states. This is consistent with what already observed in the strength indicators of superexchange interactions (cf. Fig.\ \ref{figufunction} (a)). In addition, the decreasing trends in the $J$ parameters and thus the Curie temperatures are consistent with the trends observed in the strength indicators of the more dominant FM superexchange interactions. 


\begin{figure}[t]
\subfigure[]{\includegraphics[width= 6.5 cm]{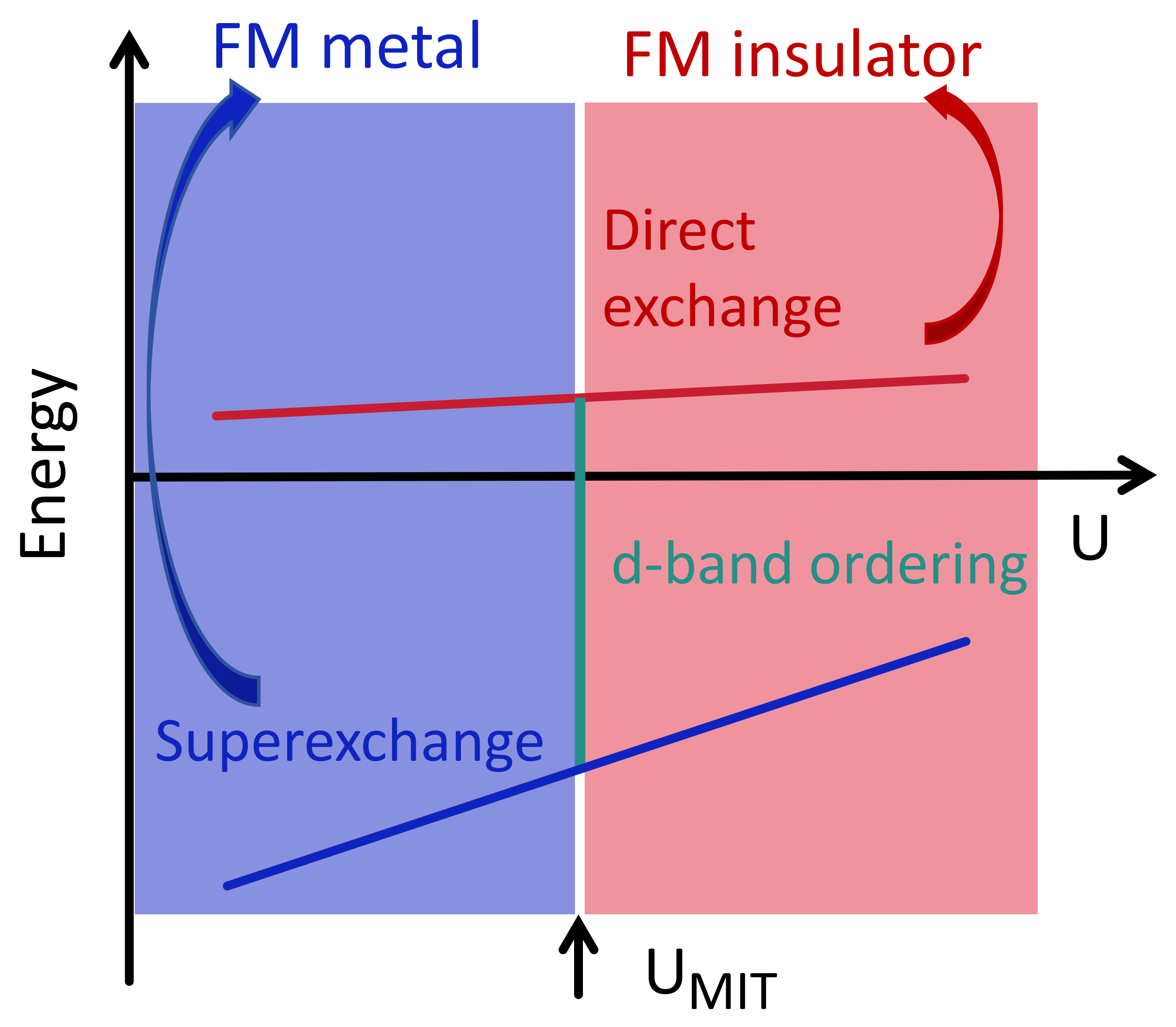}  }  
     	\caption{Schematic interrelations between the exchange interactions and the Mott metal-insulator transitions in the FM $1T$-Fe$X_2$ ($X$=Cl, Br, I) monolayers. \label{figending}}   
    \end{figure}

Experimentally, pressure can be used to induce insulator-metal transition in 2D FM monolayer CrGeTe$_3$ and the Curie temperatures are  transformed from 66 K to 250 K \cite{Bhoi2021, Sun2018, Lin2018}. Similar behaviors are experimentally observed in CrSiTe$_3$ flakes and VI$_3$ as well \cite{Zhang2021nl, Valenta2021}. These facts can be partly interpreted by the findings that  stronger FM superexchange interaction in the metallic states when compared with the insulating states in the vicinity of the MITs.

Exemplified by the FM $1T$-Fe$X_2$ ($X$=Cl, Br, I) monolayers, the interrelations between the Mott metal-insulator transitions and the exchange interaction strength in 2D FM materials have been clarified (cf. Fig.\ \ref{figending}). In the vicinity of the Mott transition points, the FM superexchange interactions counteract with the direct exchange interactions and spin-down $d$-band ordering in the FM SC states and therefore the FM superexchange interactions have a dominant role in the MITs. As a consequence, the enhanced FM superexchange interactions lead to higher Curie temperatures when crossover from the insulator to the metal sides. 

The generality of the presented physical picture is discussed as follows. In the context of superexchange interactions, the hybridizations between the anion-$p$ and cation-$e_g$ orbitals with larger orbital overlaps are stronger than the hybridizations between the anion-$p$ and cation-$t_{2g}$ orbitals \cite{Kanamori1959}. For 2D FM materials, the electronic bands derived mostly from the one or more empty cation-$e_g$ orbitals are higher than the Fermi levels, whereas the anion-$p$ bands are well below the Fermi levels. Comparing with the metallic states, the finite band gaps in the semiconducting states enlarge the energy differences between the anion-$p$ and cation-$e_g$ orbitals, which can weaken the hybridizations between them and thus the  resultant FM superexchange interactions. 







 \section{Conclusions}   

In conclusion, we employ first-principles calculations to study the fundamental interrelations between the exchange interactions and the Mott metal-insulator transition in the 2D FM $1T$-Fe$X_2$ ($X$=Cl, Br, I) monolayers. The critical behaviors  of the competing superexchange interactions and the direct exchange interactions are clarified by using the quantitative strength indicators. In particular, the FM superexchange interactions and the direct exchange interactions are found to energetically favor the halfmetallic and semiconducting states, respectively. 
When crossover from the semiconducting sides to the metallic sides, the ferromagnetic exchange interactions and thus the Curie temperatures are enhanced. This work unravels the critical behaviors of the exchange interaction near the Mott metal-insulator transitions and could be useful to the device applications which simultaneously require Mott metal-insulator transitions and exchange-interaction-related magnetic properties in 2D FM materials, particularly robust magnetic ordering and higher Curie temperatures.  \\ \\

See the supplementary material for the electronic band structure, the PDOS, the integrations of the PDOS for the spin-down $d$ orbitals for the 1T-Fe$X_2$ ($X$=Cl, Br, I) monolayers.

    \section*{Acknowledgments}
The authors were financially supported by National Natural Science Foundation of China (Grant No.\ 12304088) and Taishan Scholar Project of Shandong Province (No.\ tsqnz20221130). We used the computing facilities  at National Supercomputer Center in Guangzhou.  \\

\section*{DATA AVAILABILITY}
The data that support the findings of this article are not publicly available. The data are available from the authors upon reasonable request.  

    \bibliography{phasetrans}

\end{document}